\documentclass[conference]{IEEEtran}

\usepackage{multirow}
\usepackage[pdftex]{graphicx}
\usepackage{listings}
\usepackage{xspace}
\usepackage{booktabs}
\usepackage{flushend}
\usepackage{enumitem}
\usepackage{tcolorbox}
\usepackage[normalem]{ulem}
\usepackage{amssymb}
\usepackage{pifont}
\newcommand{\cmark}{\ding{51}}
\newcommand{\xmark}{\ding{55}}
\usepackage{subcaption}

\usepackage{siunitx}

\usepackage{algorithm}
\usepackage{algpseudocode}
\algrenewcommand\alglinenumber[1]{\tiny #1:}
\usepackage{etoolbox}
\makeatletter
\newcommand*{\algrule}[1][\algorithmicindent]{%
  \makebox[#1][l]{%
    \hspace*{0.8em}
    \vrule height .75\baselineskip depth .25\baselineskip
  }
}

\newcount\ALG@printindent@tempcnta
\def\ALG@printindent{%
    \ifnum \theALG@nested>0
    \ifx\ALG@text\ALG@x@notext
    \else
    \unskip
    \ALG@printindent@tempcnta=1
    \loop
    \algrule[\csname ALG@ind@\the\ALG@printindent@tempcnta\endcsname]%
    \advance \ALG@printindent@tempcnta 1
    \ifnum \ALG@printindent@tempcnta<\numexpr\theALG@nested+1\relax
    \repeat
    \fi
    \fi
}
\patchcmd{\ALG@doentity}{\noindent\hskip\ALG@tlm}{\ALG@printindent}{}{\errmessage{failed to patch}}
\patchcmd{\ALG@doentity}{\item[]\nointerlineskip}{}{}{} 
\makeatother

\usepackage[numbers,sort]{natbib}
\usepackage[colorlinks=true,citecolor=blue,linkcolor=blue,urlcolor=blue,draft]{hyperref}

\usepackage{microtype}
\newcommand{\js}{JavaScript\xspace}
\newcommand{\webkit}{WebKit\xspace}
\newcommand{\xml}{XML\xspace}
\newcommand{\jerry}{Jerryscript\xspace}
\newcommand{\libplist}{libplist\xspace}
\newcommand{\chakra}{ChakraCore\xspace}
\newcommand{\afl}{AFL\xspace}
\newcommand{\tool}{Superion\xspace}

\begin{document}

\title{\tool: Grammar-Aware Greybox Fuzzing}

\author{
\IEEEauthorblockN{Junjie Wang$^*$, Bihuan Chen$^\dagger$, Lei Wei$^*$, Yang Liu$^*$}
\IEEEauthorblockA{$^*$Nanyang Technological University, Singapore}
\IEEEauthorblockA{$^\dagger$Fudan University, China}
}

\maketitle

\begin{abstract}
In recent years, coverage-based greybox fuzzing~has proven itself to be one of the most effective techniques for finding security bugs in practice. Particularly, American~Fuzzy~Lop~(\afl for short) is deemed to be a great success in fuzzing relatively~simple test inputs. Unfortunately, when it meets structured test~inputs such as \xml and \js, those grammar-blind trimming and mutation strategies in \afl hinder the effectiveness and efficiency.


To this end, we propose a grammar-aware coverage-based~greybox fuzzing approach to fuzz programs that process structured~inputs. Given the grammar (which~is often publicly~available)~of~test inputs, we introduce a grammar-aware trimming strategy~to~trim test inputs at the tree level using the abstract syntax trees~(ASTs) of parsed test inputs.~Further,~we introduce two grammar-aware mutation strategies (i.e., enhanced dictionary-based mutation~and tree-based mutation). Specifically, tree-based mutation~works~via replacing subtrees using the ASTs~of~parsed test inputs.~Equipped with grammar-awareness, our approach can carry the fuzzing~exploration into width and depth.



We implemented our approach as an extension to AFL,~named \tool; and evaluated the effectiveness~of~\tool on real-life large-scale programs (a \xml engine \libplist and~three~\js engines \webkit, \jerry and \chakra). Our results have demonstrated that \tool can improve the code coverage (i.e., 16.7\% and 8.8\% in line and function coverage) and~bug-finding capability (i.e., 31 new bugs, among~which we~discovered 21 new vulnerabilities with 16 CVEs assigned and 3.2K~USD bug bounty rewards received) over AFL and jsfunfuzz. We also demonstrated the effectiveness of our grammar-aware trimming and mutation.

\end{abstract}



\section{Introduction}\label{sec:intro}

Fuzzing or fuzz testing is an automated software testing~technique to feed a large amount of invalid or unexpected test~inputs to a target program in the hope of triggering~unintended~program behaviors, e.g., assertion failures, crashes, or hangs. Since its introduction in early 1990s~\cite{Miller1990}, fuzzing has become one~of the most effective techniques for finding bugs or vulnerabilities in real-world programs. It has been~successfully~applied~to~testing various applications, ranging from rendering engines and image processors to compilers and interpreters.

A fuzzer can be classified as generation-based (e.g., \cite{yang2011finding, holler2012fuzzing, Veggalam2016, wang2017skyfire}) or mutation-based (e.g., \cite{stephens2016driller, bohmecoverage, rawat2017vuzzer, Li2017}), depending~on whether test inputs are generated by the knowledge of the~input format or grammar or by modifying well-formed test~inputs. A fuzzer can also be classified as whitebox~(e.g., \cite{godefroid2008grammar, pham2016model}),~greybox (e.g., \cite{bohmecoverage, Li2017}) or blackbox (e.g.,~\cite{Miller1990, woo2013scheduling}),~depending~on~the degree of leveraging a target program's internal structure,~which reflects the tradeoffs between effectiveness and efficiency. In this paper, we focus on mutation-based greybox fuzzing.

{\bf Coverage-Based Greybox Fuzzing.} One of the most successful mutation-based greybox fuzzing techniques is coverage-based greybox fuzzing, which uses the coverage information~of each executed test input to determine the test inputs that~should be retained for further incremental fuzzing. \afl~\cite{afl} is~a~state-of-the-art coverage-based greybox fuzzer, which has discovered thousands of high-profile vulnerabilities. Thus,~without the~loss of generality, we consider \afl as the typical implementation~of coverage-based greybox fuzzing.

\begin{figure}[!t]
\centering
\includegraphics[width=0.48\textwidth]{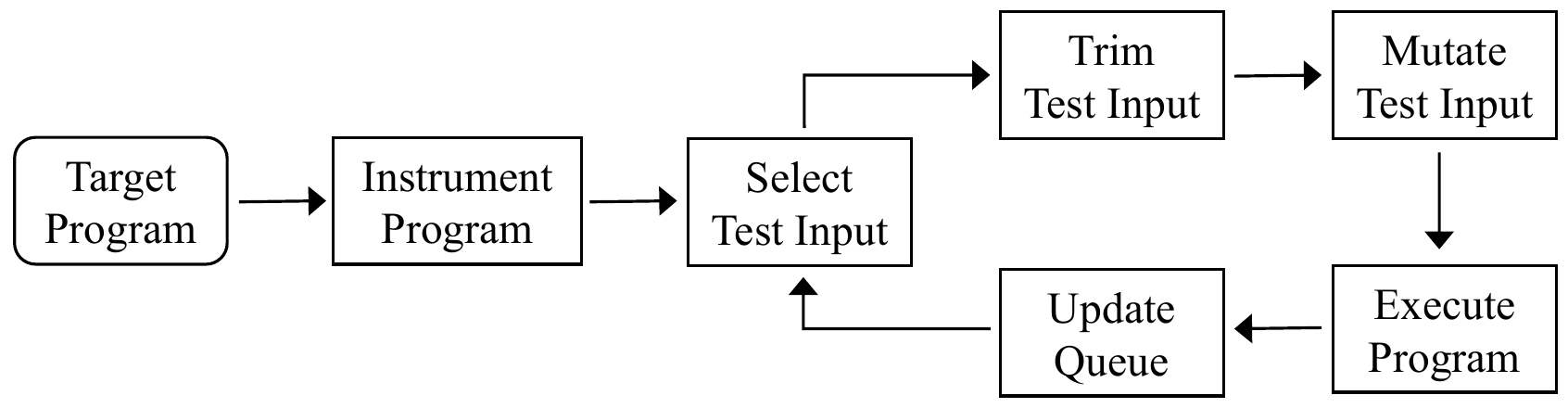}
\vspace{-5pt}
\caption{The General Workflow of \afl}
\label{fig:afl_workflow}
\end{figure}

As shown in Fig.~\ref{fig:afl_workflow}, \afl takes the target program~as~an~input, and works in two steps: instrumenting the target program and fuzzing the instrumented program. The instrumentation~step~injects code~at branch points to capture branch~(edge)~coverage together with branch hit counts (which are bucketized to small powers of two). A test input is said to have new~coverage if it either hits a new branch, or achieves a new hit count for~an already-exercised branch. The fuzzing step can be broken down into five sub-steps. Specifically, a test input is first selected~from a queue where the initial test inputs as well as the test inputs~that have new coverage are stored. Then the test input is trimmed~to the smallest size that does not change the measured behavior~of the program, as the size of test inputs has a dramatic impact~on the fuzzing efficiency. The trimmed test input is then mutated~to generate new test inputs; and the program is executed with respect to each mutated test input. Finally, the queue is updated by adding those mutated test inputs to the queue if they achieve new coverage, while the mutated test inputs that achieve no~new coverage are discarded. This fuzzing loop continues by selecting a new test input from the queue.

{\bf Challenges.} The current coverage-based greybox fuzzers~can effectively fuzz programs that process compact and unstructured inputs (e.g., images). However, some challenges arise when~they are used to target programs~that~process structured inputs (e.g., \xml and \js) that often follow specific grammars. Such programs often process the inputs in stages, i.e., syntax parsing, semantic checking, and application execution~\cite{wang2017skyfire}.

On one hand, the trimming strategies (e.g., removal of~chunks of data) in \afl are grammar-blind, and hence can easily~violate the grammar or destroy the input structure. As a result,~most~test inputs in the queue cannot be effectively trimmed to keep~them syntax-valid. This is especially the case~when the target~program can process a part of a test input (triggering coverage) but errors out on the remaining part. This will greatly affect~the~efficiency of \afl because it needs~to~spend more time on fuzzing~the test inputs whose structures are destroyed, but only finds parsing errors and gets stuck at the syntax parsing stage,~which heavily limits the capability of fuzzers in finding deep bugs.


On the other hand, the mutation strategies (e.g., bit flipping) in \afl~are grammar-blind, and hence most of the mutated~test inputs fail~to pass syntax parsing and are rejected~at~an~early stage~of~processing. As a result, it is difficult for \afl to achieve large-step mutations. For example, it is very difficult to obtain {\it Content-Length: -1} from mutating {\it Set-Cookie: FOO=BAR} via small-step bit flipping mutations~\cite{afldictionary}. Meanwhile, \afl~needs to spend a large amount of time struggling~with syntax~correctness, while only finding~parsing errors. Therefore, the effectiveness of \afl~to~find deep bugs is heavily limited~for programs that process structured inputs.

{\bf The Proposed Approach.} To address the challenges,~we~propose a new grammar-aware coverage-based greybox fuzzing~approach for programs that process structured inputs.~We~also~implement the proposed approach as an extension to \afl,~named \tool\footnote{\tool is an Autobot combiners in the cartoon \textit{The Transformers}.}. Our approach takes as inputs a {\it target program}~and~a {\it grammar} of the test~inputs that is often publicly available.~Based on the grammar, we parse each test input~into~an~abstract~syntax tree (AST). Using ASTs, we introduce a grammar-aware~trimming strategy that can effectively trim test inputs~while~keeping the input structure valid. This is realized~by~iteratively~removing each~subtree in the AST of a test input and observing~coverage differences. Moreover, we propose two grammar-aware mutation strategies that can quickly carry the fuzzing exploration~beyond syntax parsing. We first enhance \afl's dictionary-based mutation strategy by inserting/overwriting tokens~in~a grammar-aware manner, and then propose a tree-based mutation strategy that replaces one subtree in the AST of a test input~with the subtree from itself or another test input~in~the queue. 

\begin{figure}[!t]
\centering
\includegraphics[width=0.48\textwidth]{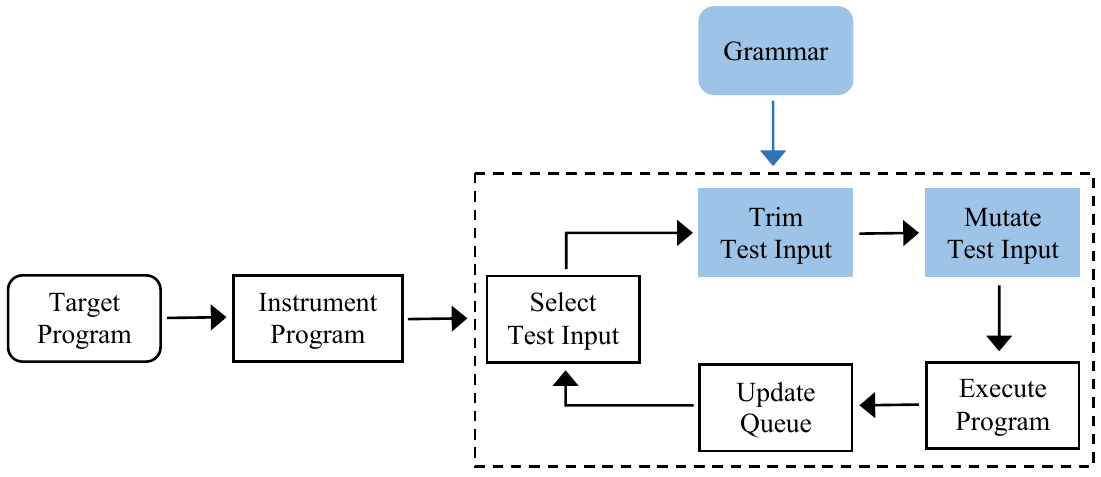}
\vspace{-5pt}
\caption{The General Workflow of \tool with the Highlighted Differences from AFL (see Fig.~\ref{fig:afl_workflow})}
\label{fig:overview}
\end{figure}

To evaluate the effectiveness of \tool, we conducted~experiments on one \xml engine \libplist  and~three~\js~engines \webkit, \jerry and \chakra. We compared~our approach with AFL with respect to the code coverage and bug-finding capability. The results have demonstrated that \tool can effectively improve the code coverage over AFL by 16.7\% in line coverage and 8.8\% in function coverage;~and \tool can~significantly improve the bug-finding capability over AFL by finding 31 new bugs (among which six were found by AFL). Among these bugs, 21 new vulnerabilities were discovered~with 16 CVEs assigned; and we received 3.2K~USD bug bounty rewards. Besides, we compared \tool with jsfunfuzz \cite{ruderman2007introducing}, which is a successful fuzzer specifically designed for \js. However, it failed to find any new bugs. Moreover, we have demonstrated that our grammar-aware trimming strategy can effectively trim test inputs while keeping them syntax-valid;~and our grammar-aware mutation strategies can effectively generate new test inputs that can trigger new coverage.

{\bf Contributions.} The contributions of this work are:
\begin{itemize}[leftmargin=*]
\item We proposed a novel grammar-aware coverage-based greybox fuzzing approach for programs that process structured inputs, which complements existing coverage-based greybox fuzzers.
\item We implemented our approach and made it open-source\footnote{https://github.com/zhunki/gramFuzz},~and conducted experiments to~demonstrate its effectiveness.
\item We found 31 new bugs, among which we found 21 new~vulnerabilities with 16 CVEs assigned and received 3.2K USD bug bounty rewards.
\end{itemize}


\section{Our Approach}\label{sec:approach}

To address the challenges of coverage-based greybox fuzzing (Section~\ref{sec:intro}), we propose a novel grammar-aware coverage-based greybox fuzzing approach, which targets programs~that~process structured inputs. We implement the approach as an extension~to \afl~\cite{afl}, named \tool. Fig.~\ref{fig:overview} introduces the workflow~of \tool, and highlights the differences from \afl (see~Fig.~\ref{fig:afl_workflow}). In particular, a context-free grammar of the test inputs~is~needed, which is often publicly available (e.g., in ANTLR's community \cite{grammarlist}). We introduce a grammar-aware trimming strategy (Section \ref{subsec:trim}) and two grammar-aware mutation strategies (Section \ref{subsec:mutation}) with the purpose of making \afl grammar-aware.


\subsection{Grammar-Aware Trimming Strategy}\label{subsec:trim}

The built-in trimming strategy in \afl is grammar-blind,~and treats a test input as chunks of data. Basically, it first~divides~the test input to be trimmed into chunks of $len/n$ bytes where $len$ is the length of the test inputs in bytes, and then~tries to remove each chunk sequentially. If the coverage remains~the same after~the removal of a chunk, this chunk is trimmed. Note that $n$ starts at 16 and increments by a power of two~up~to~1024. This strategy is very effective for unstructured inputs. However, it cannot effectively prune structured inputs~while~keeping~them syntax-valid, possibly making \afl stuck in the fuzzing~exploration of syntax parsing without finding deep bugs.

\begin{figure}[!t]
\tiny
\centering
\begin{lstlisting}[frame=single,language=xml,basicstyle=\ttfamily,escapeinside={@}{@}]
<?xm@\sout{l \textbf{versio}}\textbf{n}@="1.0" encoding="UTF-8"?>
<plist version="1.0">
<dict>
	<key>Some ASCII string</key>
	<string></string>
	<data>
	</data>
</@\sout{dict>}@
@\sout{</plis}@t>
\end{lstlisting}
\vspace{-5pt}
\caption{An Example of AFL's Built-In Trimming}
\label{fig:trimafl}
\end{figure}

{\bf Example.} Fig.~\ref{fig:trimafl} gives an example of AFL's built-in trimming on an XML test input with respect to \libplist (an XML engine), where ``l versio'' and ``dict$>$ $<$/plis''~are trimmed (highlighted by strikethrough). The trimmed test input is syntax-invalid,~but has the same coverage with the original test input  due to the~gap between implementation of \libplist and grammar specification. Hence,~the trimmed test input is used for further fuzzing even though its grammar is destroyed by AFL's built-in trimming.

\begin{algorithm}[!t]
\scriptsize
\caption{Grammar-Aware Trimming}
\label{alg:trim}
\begin{algorithmic}[1]
\Require the test input to be trimmed $in$, the grammar $G$
\Ensure the trimmed test input $ret$
\While {$true$}
\State parse $in$ according to $G$ into an AST $tree$ \label{alg1:2}
\If {there are any parsing errors} \label{alg1:3}
\Return built-in-trimming ($in$)
\EndIf \label{alg1:5}
\For {each subtree $n$ in $tree$}\label{alg1:6}
\State $ret = $ remove $n$ from $tree$\label{alg1:7}
\State run the target program against $ret$
\If {coverage remains the same} \label{alg1:9}
\State $in = ret$
\State {\bf break} \label{alg1:11}
\Else \label{alg1:12}
\State add $n$ back to $tree$\label{alg1:13}
\EndIf \label{alg1:14}
\If {$n$ is the last subtree in $tree$}\label{alg1:15}
\Return $ret$ \label{alg1:16}
\EndIf
\EndFor\label{alg1:18}
\EndWhile
\end{algorithmic}
\end{algorithm}

To ensure the syntax-validity of trimmed test inputs,~we~propose a grammar-aware trimming strategy, whose procedure~is given in Algorithm \ref{alg:trim}. It first parses the test input to~be~trimmed $in$ according to the grammar $G$ into an AST $tree$ (Line~\ref{alg1:2}).~If any parsing errors occur (as $in$'s structure may be destroyed~by mutations), then it uses \afl's built-in trimming strategy~rather than directly discarding it (Line~\ref{alg1:3}--\ref{alg1:5}); otherwise, it attempts~to trim a subtree~$n$~from $tree$ (Line~\ref{alg1:6}--\ref{alg1:7}). If the coverage~is~different after $n$ is trimmed, then $n$ cannot be trimmed (Line~\ref{alg1:12}--\ref{alg1:14}), and it tries to trim~next subtree; otherwise, $n$ is trimmed, and it re-parses the remaining test input (Line~\ref{alg1:9}--\ref{alg1:11}), and then repeats the procedure until no subtree can be trimmed (Line~\ref{alg1:15}--\ref{alg1:16}). Thus, we resort to AFL's built-in trimming only when our~tree-based trimming is  not applicable. This is because sometimes invalidity is also useful.


\begin{figure}[!t]
\tiny
\centering
\begin{lstlisting}[frame=single,language=java,basicstyle=\ttfamily,escapeinside={@}{@}]
...
try{eval("M:if(([15,16,17,18].some(this.unwatch(\"x\"),(([window if([[]])])[this.prototype])))) else{true;return null;}");} catch(ex){}
try{eval("M:while((null >=\"\")&&0){/a/gi}");} catch(ex){}
try{eval("\nbreak M;\n");} catch(ex){}
try{eval("L:if((window[(1.2e3.x::y)]).x) return null; else if((uneval(window))++.propertyIsEnumerable(\"x\")){CollectGarbage()}");} catch(ex){}
@\sout{\textbf{try}\{eval("/*for..in*/for(var x in (((\{\}).hasOwnProperty}@
     @\sout{)([,,].hasOwnProperty($\setminus$"x$\setminus$"))))/*for..in*/ M:for(var}@
     @\sout{[window, y] =(-1) in this) [1,2,3,4].slice");\} \textbf{catch}}@
     @\sout{(ex)\{\}}@
try{eval("if(\"\"){}else if(x4) {null;}");} catch(ex){}
try{eval("{}");} catch(ex){}
try{eval("for(var x = x in x - /x/ ){}");} catch(ex){}
try{eval("if((uneval(x, x))) var x = false; else if((null\n.unwatch(\"x\"))) throw window; else {} return 3;");}catch(ex){}
...
\end{lstlisting}
\vspace{-5pt}
\caption{An Example of Grammar-Aware Trimming}
\label{fig:trim}
\end{figure}

{\bf Example.} Fig.~\ref{fig:trim} shows an example of our trimming~strategy on a JavaScript test input, where a complete {\tt try-catch}~statement (highlighted by strikethrough) is trimmed~without introducing any coverage~differences. However, it is almost~impossible for \afl's built-in trimming strategy to prune such~a complete statement.


\subsection{Grammar-Aware Mutation Strategies}\label{subsec:mutation}

The default mutation strategies (e.g., bit flipping or token~insertion) in \afl are too fine-grained and grammar-blind~to~keep the input structure following the underlying grammar. Therefore, we propose two grammar-aware mutation strategies to improve the mutation effectiveness on triggering new program behaviors.

\subsubsection{Enhanced Dictionary-Based Mutation}



Dictionary-based mutation~\cite{afldictionary} was introduced to make up for the grammar-blind nature of \afl. The dictionary is referred as a list~of~basic~syntax tokens (e.g., reserved keywords) which can be provided~by~users or automatically identified by \afl. Every token is inserted~between every two bytes of the test input to be mutated,~or~written over every byte sequence of the same length of the token. Such mutations can generate syntax-valid test inputs but is inefficient as most of the generated inputs have destroyed structure.

\begin{algorithm}[!t]
\scriptsize
\caption{Dictionary-Based Mutation}
\label{alg:dictionary}
\begin{algorithmic}[1]
\Require the test input $in$, the dictionary $D$
\Ensure the set of mutated test inputs $T$
\State $T = \emptyset$
\State $l = $ the length of $in$
\For { $i = 0$; $i < l$;} \label{alg2:3}
\State $j = i + 1$
\State $curr = $ *(u8*)($in$'s address + $i$)~~// current byte of $in$
\State $next = $ *(u8*)($in$'s address + $j$)~~// next byte of $in$
\While {$j <  l$ \&\& $curr$ and $next$ are alphabet or digit}
\State $j = j + 1$
\State $next =$ *(u8*)($in$'s address + $j$)
\EndWhile \label{alg2:10}
\For {each token $d$ in $D$}\label{alg2:11}
\State insert $d$ at $i$ of $in$ / overwrite $i$ to $j$ of $in$ with $d$
\State $T$ = $T \cup \{in\}$
\EndFor\label{alg2:14}
\State $i$ = $j$
\EndFor
\end{algorithmic}
\end{algorithm}

Therefore, we propose the enhanced dictionary-based mutation as shown in Algorithm~\ref{alg:dictionary}. This algorithm leverages~the~key fact that the tokens (e.g., variable names, function~names,~or~reserved keywords) in a structured test input normally only~consist of alphabets~or digits. Hence, it first locates the token~boundaries in~a~test~input by iteratively checking whether the current and next byte are both alphabet~or~digit (Line~\ref{alg2:3}--\ref{alg2:10}). Then it inserts each token in the dictionary~to~each located boundary, which avoids the insertion between consecutive sequence of alphabets and digits and thus greatly decreases the number~of token insertions (Line~\ref{alg2:11}--\ref{alg2:14}). Similarly, it writes each token in the dictionary over the content between every~two located boundaries, which also greatly decreases the number~of token overwrites. Such token insertions and overwrites~not~only maintains the structure of mutated test inputs but also decreases the number of mutated test inputs, hence greatly improving~the effectiveness and efficiency of dictionary-based mutation.

\begin{figure}[!t]
\centering
\begin{subfigure}[b]{0.23\textwidth}
\centering
\includegraphics[width=0.95\textwidth]{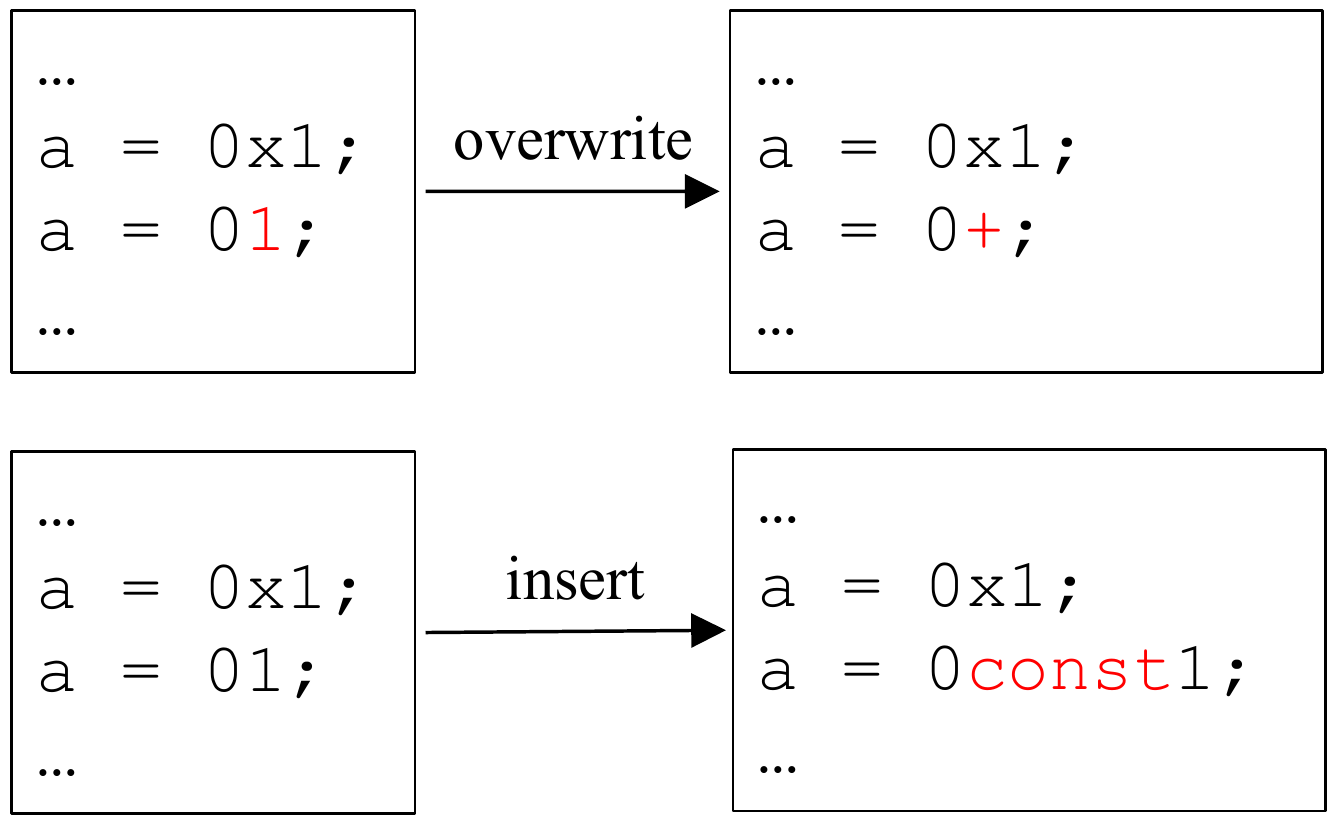}
\caption{Original}\label{fig:sample1a}
\end{subfigure}%
~~~
\begin{subfigure}[b]{0.23\textwidth}
\centering
\includegraphics[width=0.95\textwidth]{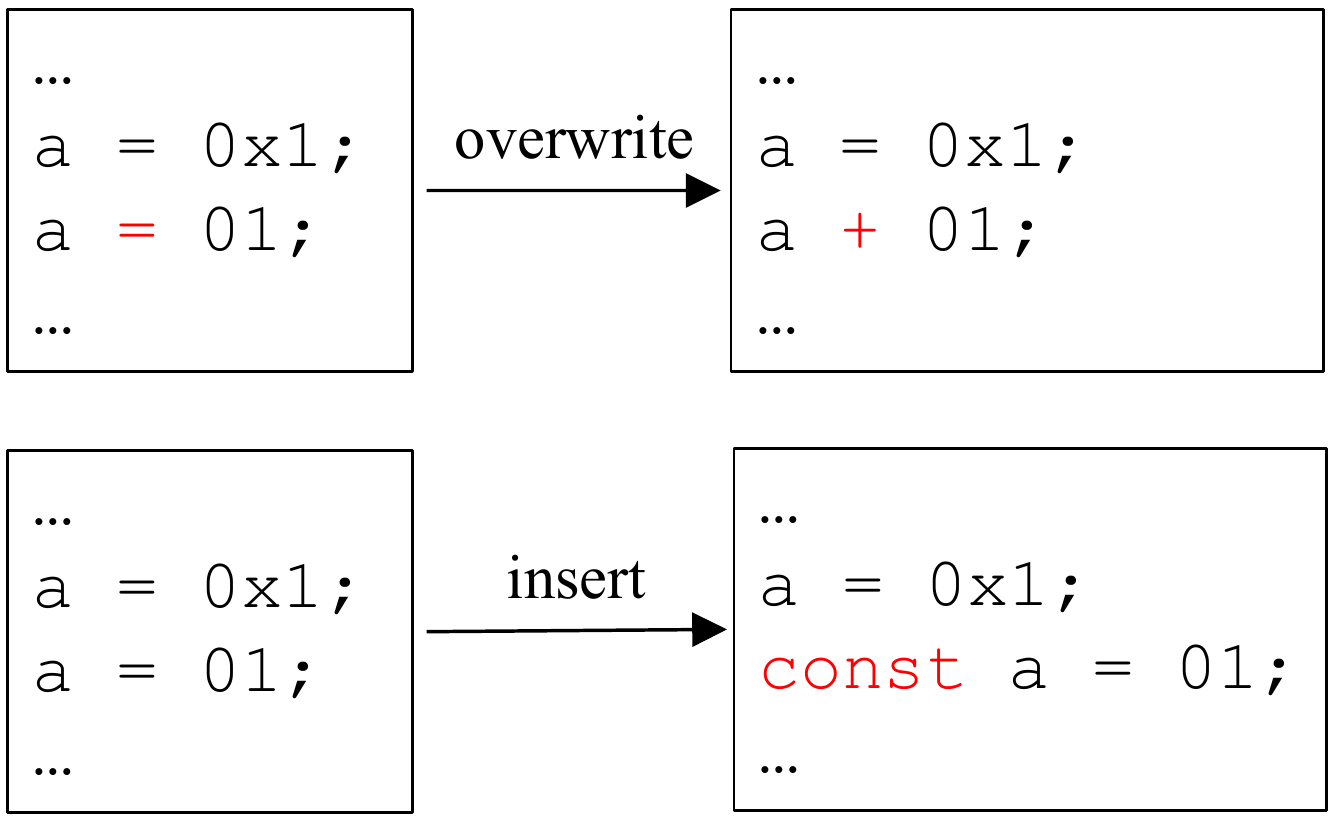}
\caption{Enhanced}
\label{fig:sample1b}
\end{subfigure}
\caption{An Example of Dictionary-Based Mutation}
\label{fig:sample1}
\end{figure}

{\bf Example.} Fig.~\ref{fig:sample1} illustrates the difference between the~original and enhanced dictionary-based mutation. In the original~one, {\tt 01} is not treated as a whole,~and thus {\tt 1} can be overwritten~by {\tt +} and {\tt const} can be inserted between {\tt 0} and {\tt 1}, which destroys the structure without introducing any new coverage.~In~the~enhanced one, {\tt 01} is identified as a whole, and hence the mutated test inputs in Fig.~\ref{fig:sample1a}~will~not~be~produced.~Instead,~it~can~generate the mutated test inputs in Fig.~\ref{fig:sample1b} more efficiently, which are taken from our experiments and both lead to new coverage.

\subsubsection{Tree-Based Mutation}

Dictionary-based mutation is aware of the underlying grammar in an implicit way. To be explicitly aware of the grammar and thus producing syntax-valid~test~inputs, we utilize the grammar knowledge and~design~a~tree-based mutation, which works at the level of ASTs. Different~from~the tokens used in dictionary-based mutation, AST actually models a test input as objects with named properties, and is designed~to represent all the information about a test input. Thus, ASTs~provide a suitable granularity for a fuzzer to mutate test inputs.



Algorithm~\ref{alg:mutation} shows the procedure of our tree-based mutation. It takes as inputs a test input $tar$ to be mutated, the grammar~$G$, and a test input $pro$ that is randomly chosen from the queue.~It first parses $tar$ according to $G$ into an AST $tar\_tree$; and~if any parsing errors occur, $tar$ is a syntax-invalid test input and we do not apply tree-based mutation to $tar$ (Line~\ref{alg3:3}--\ref{alg3:6}).~If~no error occurs, it traverses $tar\_tree$, and stores each subtree in a set $S$ (Line~\ref{alg3:7}--\ref{alg3:9}). Then it parses $pro$ into an AST $pro\_tree$,~and stores each subtree of $pro\_tree$ in $S$ if there is no parsing~error (Line~\ref{alg3:10}--\ref{alg3:15}). Here $S$ serves as the~content~provider of mutation. Then, for each subtree $n$ in $tar\_tree$, it replaces $n$ with~each~of the subtree $s$ in $S$ to generate a new mutated test input (Line~\ref{alg3:16}--\ref{alg3:21}). Finally, it returns the set of mutated test inputs.

The size of this returned set can be the multiplication~of~the number of subtrees in $tar\_tree$ and the~number~of~subtrees~in $tar\_tree$ and $pro\_tree$, which could be very large.~As~an~example, our tree-based mutation on $tar$ and $pro$~whose~number~of subtrees is respectively 100 and 500 will generate $100 \times (100 +$ $500)= 60,000$ test inputs. This will add burden to the program execution step during fuzzing, making fuzzing less efficient.~To relieve the burden, we design~three~heuristics~to~reduce~the~number of mutated test inputs. For clarity, we do not elaborate~these heuristics in Algorithm~\ref{alg:mutation}, but only show where they are applied.

\begin{algorithm}[!t]
\scriptsize
\caption{Tree-Based Mutation}
\label{alg:mutation}
\begin{algorithmic}[1]
\Require the test input $tar$, the grammar $G$, the test input $pro$
\Ensure the set of mutated test inputs $T$
\State $T = \emptyset$
\State $S = \emptyset$~~// the set of subtrees in $tar$ and $pro$
\State parse $tar$ according to $G$ into an AST $tar\_tree$~~// Heuristic~1\label{alg3:3}
\If {there are any parsing errors}
\Return
\EndIf \label{alg3:6}
\For {each subtree $n$ in $tar\_tree$}~~// Heuristic 3 \label{alg3:7}
\State $S = S \cup \{n\}$
\EndFor\label{alg3:9}
\State parse $pro$ according to $G$ into an AST $pro\_tree$~~// Heuristic 1 \label{alg3:10}
\If {there is no parsing error}
\For {each subtree $n$ in $pro\_tree$}~~// Heuristic 3 \label{alg3:12}
\State $S = S \cup \{n\}$
\EndFor
\EndIf \label{alg3:15}
\For{each subtree $n$ in $tar\_tree$}~~// Heuristic 2 \label{alg3:16}
\For{each subtree $s$ in $S$}
\State $ret =$ replace $n$ in $tar\_tree$'s copy with $s$
\State $T = T \cup \{ret\}$
\EndFor
\EndFor\label{alg3:21}
\Return $T$
\end{algorithmic}
\end{algorithm}

\begin{figure*}[htbp]
\centering
\includegraphics[width=0.95\textwidth]{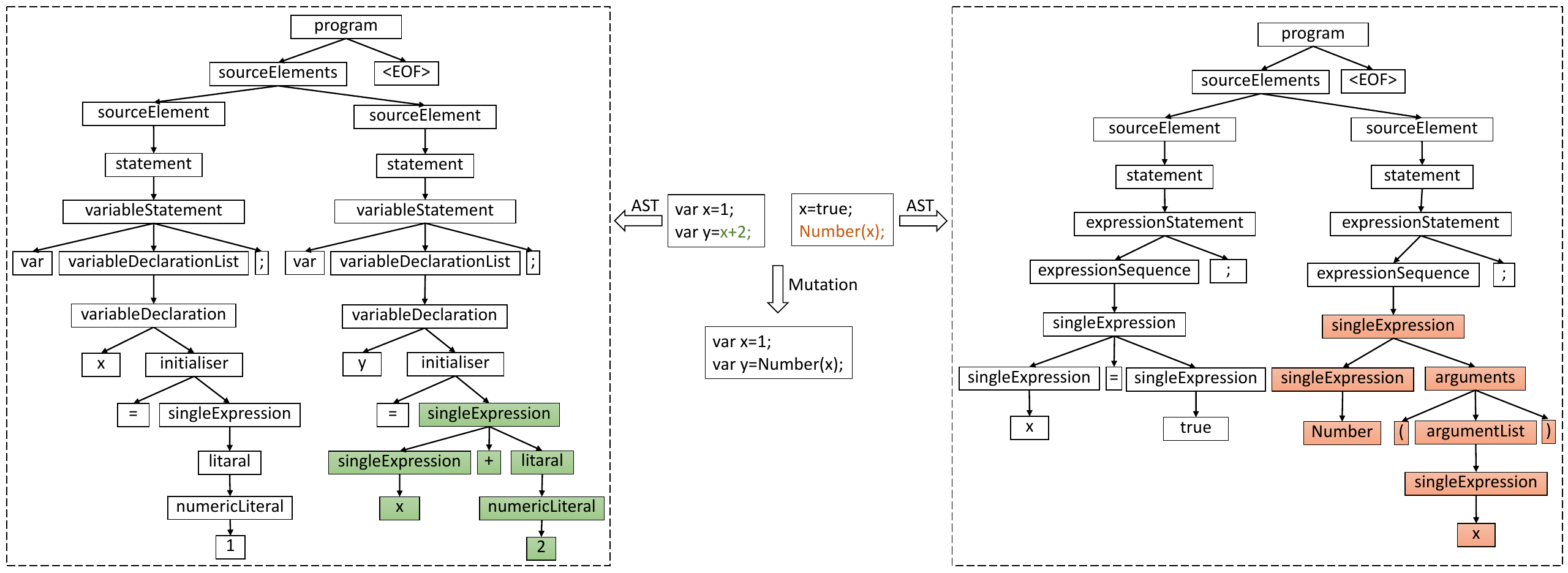}
\caption{An Example of Tree-Based Mutation}
\label{fig:sample2}
\end{figure*}

\begin{itemize}[leftmargin=*]
\item{\bf Heuristic 1: Restricting the size of test inputs.} We~limit~the size of test inputs (i.e., $tar$ and $pro$ in Algorithm~\ref{alg:mutation})~as~10,000 bytes long (Line~\ref{alg3:3} and \ref{alg3:10}). Hence we do not apply~tree-based mutation to $tar$ if $tar$ is more than 10,000 bytes~long;~and~we do not use subtrees of $pro$ as the content provider~of~mutation if $pro$~is more than 10,000 bytes long.~The reasons~are~that,~a larger test input usually needs a larger number of mutations; more memory is required to store the AST of a larger~test~input; and a larger test input often has a slower execution~speed.

\item{\bf Heuristic 2: Restricting the number of mutations.} If there are more than 10,000 subtrees in $tar$ and $pro$, we~randomly select 10,000 from all subtrees in $S$ as the content provider of mutation (Line~\ref{alg3:16}). Thus, we keep the number of mutations on each test input in the queue under 10,000~to make sure that each test input in the queue has the chance~to get mutated.

\item{\bf Heuristic 3: Restricting the size of subtrees.} We limit the size of subtrees (i.e., each subtree in $S$ in Algorithm~\ref{alg:mutation})~as~200 bytes long (Line~\ref{alg3:7} and \ref{alg3:12}). Thus we do not use~the~subtrees~of $tar$ and $pro$ as the content provider of mutation if the subtree is more than 200 bytes long. Notice that 200 bytes are long enough to include complex statements.
\end{itemize}

The threshold values in these heuristics were~empirically established as good ones.





{\bf Example.} Fig.~\ref{fig:sample2} shows an example of our tree-based~mutation. The left-side is the AST of the test input~to~be~mutated~(i.e., $tar$ in Algorithm~\ref{alg:mutation}), and the right-side is the~AST~of~the~test~input that provides the content of mutation (i.e., $pro$ in Algorithm \ref{alg:mutation}). Here the subtree corresponding to the expression $x+2$ in~$tar$ is replaced with the subtree corresponding to the expression $Number(x)$ in $pro$, resulting in a new test input.



\section{Evaluation}\label{sec:experiment}


We implemented \tool in 3,372 lines of C/C++~code by extending AFL~\cite{afl}. Particularly, given the grammar~of~test~inputs, we adopted~ANTLR~4~\cite{parr2013definitive} to generate the lexer and~parser, and used ANTLR~4 C++ runtime to parse~test inputs and realize our trimming and mutation strategies. Hence, our approach is general and easily adoptable for other structured test inputs. 

\subsection{Evaluation Setup}

To evaluate the effectiveness and generality of our approach, we selected two target languages and four target programs, and compared our approach with \afl~\cite{afl}~with~respect~to~the~bug-finding capability and code coverage.

\begin{table}[!t]
\centering
\small
\caption{Target Languages and Their Structure and Samples}\label{tab:languages}
\def\arraystretch{1.25}
\setlength{\tabcolsep}{0.5em}
\vspace{-5pt}
\begin{tabular}{cccc}
\hline
Language &\# Symbols & Structure Level & \# Samples \\\hline
\xml  & 8             & Weak    & 9,467 (534) \\
\js   & 98            & Strong  & 20,845 (2,569) \\
\hline
\end{tabular}
\end{table}

{\bf Target Languages.} We chose \xml and \js~as~the~target languages with different structure level. Their grammars~are all publicly available in ANTLR's~community~\cite{grammarlist}. In particular, \xml is a widely-used markup language, which defines~a~set~of rules for encoding documents. It has been widely used in a variety of applications. As shown in the~second~column~of~Table~\ref{tab:languages}, the \xml grammar only contains~eight~symbols.~Thus,~\xml can be considered~to~be weakly-structured. On the other hand, \js is an interpreted programming~language,~which is employed by most~websites and supported by all modern web browsers. The \js grammar contains 98 symbols, and thus its structure level can be regarded as strong.


As indicated by the last column of Table~\ref{tab:languages}, we crawled~9,467 \xml samples from the Internet, and 20,845 \js samples from the test inputs of the two open-source \js engines \webkit and \jerry. They were used as the initial test inputs (i.e.,~seeds)~for fuzzing. As suggested by \afl, {\it afl-cmin} should be used~to~identify the set of functionally distinct seeds that exercise different code paths in the target program when a large amount of seeds are available. Therefore, we used {\it afl-cmin} on the samples,~and identified 534 and 2,569 distinct \xml and \js samples as the seeds for fuzzing, as shown in the parentheses in the last column of Table~\ref{tab:languages}. Notice that, before fuzzing, we pre-processed the \js samples by removing all the comments as comments, especially multi-line comments, account for a considerable percentage of waste of mutation.

{\bf Target Programs.} We selected one open-source \xml~engine \libplist and three open-source \js engines \webkit, \jerry and \chakra as the programs for fuzzing. The first~four columns of Table~\ref{tab:targets} list the program details,~including the version, the number of lines of code, and~the~number~of~functions. Particularly, \libplist is a small portable C library~to~handle Apple Property List format files in binary or \xml. It is widely used on iOS and Mac OS. \webkit is a cross-platform web browser engine. It powers Safari, iBooks and App~Store, and various Mac OS, iOS and Linux applications. \jerry~is~a lightweight \js engine for Internet of Things, intended~to run on a very constrained devices. \chakra is the core part of the Chakra Javascript engine that powers Microsoft Edge. We chose these programs because they are security-critical and widely-fuzzed. Thus, finding bugs in them are significant.


\begin{table}[!t]
\centering
\small
\caption{Target Programs and Their Fuzzing Configuration}\label{tab:targets}
\def\arraystretch{1.25}
\setlength{\tabcolsep}{0.4em}
\vspace{-5pt}
\begin{tabular}{cccccc}
\hline
Program     & Version  & \# Lines   & \# Func.  & Coverage  & Timespan \\\hline
\libplist   & 1.12     & 3,317      & 316       & Edge      & 3 months \\
\webkit     & 602.3.12 & 151,807    & 60,340    & Block     & 3 months \\
\jerry      & 1.0      & 19,963     & 1,100     & Edge      & 3 months \\
\chakra      & 1.10.1      & 236,881     & 74,132     & Block      & 3 months \\
\hline
\end{tabular}
\end{table}

As shown in the fifth column of Table~\ref{tab:targets},~we~used~edge~coverage for \libplist and \jerry during fuzzing, but block~coverage for~others due to non-determinism (i.e., different~executions of a test input lead to different coverage). Besides,~we~excluded the non-deterministic code in \webkit and \chakra from instrumentation, following the technique in kAFL~\cite{schumilo2017kafl}.

At the time of writing, we have fuzzed these programs~for about three months. For \libplist and \jerry, we have~completed more than 100 cycles of fuzzing. For~\webkit and~\chakra, due to their large size, we have not finished one cycle yet. Here a cycle means the fuzzer went over all~the interesting test inputs (triggering new coverage) discovered so far, fuzzed them, and looped back to the very beginning.


\textbf{Research Questions.} Using the previous evaluation setup, we aim to answer the following five research questions.

\begin{itemize}[leftmargin=*]
\item \textbf{RQ1:} How is the bug-finding capability of \tool?
\item \textbf{RQ2:} How is the code coverage of \tool?
\item \textbf{RQ3:} How effective is our grammar-aware trimming?
\item \textbf{RQ4:} How effective is our grammar-aware mutation?
\item \textbf{RQ5:} What is the performance overhead of \tool?
\end{itemize}

We conducted all the experiments on machines with 28 Intel Xeon CPU E5-2697v3 cores and  64GB memory, running 64-bit Ubuntu 16.04 as the operating system.

\subsection{Discovered Bugs and Vulnerabilities (RQ1)}\label{sec_bugs}

\begin{table}[!t]
\centering
\scriptsize
\caption{Unique Bugs Discovered by \tool}\label{tab:bugs}
\def\arraystretch{1.25}
\setlength{\tabcolsep}{0.5em}
\vspace{-5pt}
\begin{tabular}{ccccc}
\hline
Program                    & Bug & Type & \afl & jsfunfuzz \\\hline
\multirow{11}{*}{\libplist}& CVE-2017-5545   & Buffer Overflow &\xmark &N/A \\ 
                           & CVE-2017-5834   & Buffer Overflow & \cmark &N/A \\ 
                           & CVE-2017-5835   & Memory Corruption &\cmark &N/A \\ 
                           & CVE-2017-6435   & Memory Corruption & \xmark &N/A \\ 
                           & CVE-2017-6436  & Memory Corruption & \xmark &N/A \\ 
                           & CVE-2017-6437   & Buffer Overflow &  \cmark &N/A \\ 
                           & CVE-2017-6438   & Buffer Overflow &  \cmark &N/A \\ 
                           & CVE-2017-6439   & Buffer Overflow & \xmark &N/A \\ 
                           & CVE-2017-6440   & Memory Corruption & \xmark &N/A \\ 
                           & Bug-90          & Assertion Failure & \xmark &N/A \\ 
                           & CVE-2017-7440   & Integer Overflow & \cmark &N/A \\\hline 
\multirow{12}{*}{\webkit}  & CVE-2017-7095   & Arbitrary Access & \xmark & \xmark \\ 
                           & CVE-2017-7102  & Arbitrary Access & \xmark & \xmark \\ 
                           & CVE-2017-7107  & Integer Overflow & \xmark & \xmark \\ 
                           & Bug-188694   & Buffer Overflow & \xmark & \xmark \\ %
                           & Bug-188298   & Use-After-Free & \xmark & \xmark \\ 
                           & Bug-188917   & Assertion Failure & \xmark & \xmark \\ 
                           & Bug-170989   & Assertion Failure & \xmark & \xmark \\ 
                           & Bug-170990   & Assertion Failure & \xmark & \xmark \\ 
                           & Bug-172346   & Null Pointer Deref & \xmark & \xmark \\ 
                           & Bug-172957   & Null Pointer Deref & \xmark & \xmark \\ 
                           & Bug-172963   & Buffer Overflow & \xmark & \xmark \\ 
                           & Bug-173305   & Assertion Failure & \xmark & \xmark \\ 
                           & Bug-173819   & Assertion Failure & \xmark & \xmark \\\hline 
\multirow{4}{*}{\jerry}    & CVE-2017-18212 & Buffer Overflow & \xmark & N/A \\
& CVE-2018-11418 & Buffer Overflow & \cmark & N/A \\
& CVE-2018-11419 & Buffer Overflow & \xmark & N/A \\
& Bug-2238 & Buffer Overflow & \xmark & N/A \\
\hline
\multirow{3}{*}{\chakra}    & Bug-5534 & Buffer Overflow & \xmark & \xmark \\
& Bug-5533 & Null Pointer Deref & \xmark & \xmark \\
& Bug-5532 & Null Pointer Deref & \xmark & \xmark \\
\hline
\end{tabular}
\end{table}

Table~\ref{tab:bugs} lists the unique bugs discovered in the~four~programs by \tool. In \libplist, we discovered~11 new bugs,~from which we found~10~new~vulnerabilities with CVE identifiers assigned. 
In~\webkit, 13 new bugs were~found; and 6 of them were~vulnerabilities with 3 CVE identifiers~assigned, while others are pending for advisories. It is worth mentioning~that these bugs obtained high appraisals, e.g.,~``{\it This bug is really~interesting}'', ``{\it Thank you for the awesome test case}'' and ``{\it This~bug~has existed for a long time. A quick~look through blame would say for 4-5 years or so}''. In \jerry, we~found 4~previously unknown bugs, from which we found 4 vulnerability with 3 CVE identifiers assigned. In \chakra, we discovered 3 new bugs, and one of them is a vulnerability. Note that we received 3.2K USD bug bounty rewards.


With respect to the type of these bugs (see the third~column of Table~\ref{tab:bugs}), 12 of them are buffer overflow,~2~of~them~are~integer overflow, 4 of them are memory corruption, 2~of~them~are~arbitrary address access, and 1 of them is use-after-free. These are all vulnerabilities. Besides, 4 of them are null pointer~dereference, and 6 of them are assertion failure. These are all denial of service bugs. All these 31 bugs have been confirmed, and 25 of them have been fixed.


\textbf{Comparison to AFL.} Among these 31 bugs, \afl only~discovered six of them (as shown in the~fourth column of Table~\ref{tab:bugs}) and did not~discover~any other new~bugs. This~demonstrates that our approach significantly improves the bug finding capability of coverage-based grey-box fuzzers, which owes~to~the grammar-awareness in \tool. Specifically, for relatively weakly-structured inputs such as \xml, \afl itself found~5 bugs, while \tool not only found all these 5 bugs, but also found 6 more bugs than \afl. Differently, for highly-structured inputs such as \js, \afl barely found bugs.~Only one bug about utf-8 encoding problem was found by \afl in \jerry. All other bugs in \js engines were actually~found~by \tool's tree-based mutation. This further demonstrates the significance of injecting grammar-awareness into coverage-based grey-box fuzzers.

\textbf{Comparison to jsfunfuzz.} We also compared \tool~with jsfunfuzz~\cite{ruderman2007introducing}, which is a successful grammar-aware fuzzer specifically designed for testing \js engines. jsfunfuzz can be used to fuzz \webkit and \chakra; but it fails~to~fuzz \jerry because its generated \js inputs have many \js features that are not supported by \jerry. After three months of fuzzing, jsfunfuzz only found hundreds~of~out-of-memory crashes in \webkit and \chakra, but failed to find any bugs (as indicated by the last column of Table~\ref{tab:bugs}). This is because jsfunfuzz uses manually-specified rules to express the grammar rules the generated inputs should satisfy. However, it is daunting, or even impossible to manually express all the required rules. Instead, \tool directly uses~the grammar automatically during trimming and mutation.

\vspace{2pt}
\noindent
\fbox{
\begin{minipage}{0.455\textwidth}
In summary, \tool can significantly improve the bug-finding capability of coverage-based grey-box fuzzers (e.g., we found 31 new bugs, among which we discovered 21 new vulnerabilities with 16 CVE identifiers assigned).
\end{minipage}
}


\subsection{Code Coverage (RQ2)}\label{sec_coverage}

As empirically studied that 1\% increase in code coverage~can increase the percentage of found bugs by 0.92\%~\cite{cmiller2008fuzz}.~Hence, apart from the bug-finding capability, we measured the code coverage achieved by fuzzing. The results are shown in Table~\ref{tab:coverage}, including the line and function coverage of the target programs. In particular, we list the coverage achieved by initial seeds, \afl and \tool.~The~coverage~was~calculated using {\it afl-cov}~\cite{afl_cov}. We were not able to calculate the coverage~for~jsfunfuzz~due~to two reasons: jsfunfuzz does not keep the \js samples executed; and jsfunfuzz is verfy efficient and executes millions of \js samples until it triggers a crash, which makes the coverage computation infeasible.

\begin{table}[!t]
\centering
\scriptsize
\caption{Code Coverage of the Target Programs}
\label{tab:coverage}
\def\arraystretch{1.25}
\setlength{\tabcolsep}{0.45em}
\vspace{-5pt}
\begin{tabular}{ccccccc}
\hline
\multirow{2}{*}{Program}   & \multicolumn{3}{c}{Line Coverage (\%)} & \multicolumn{3}{c}{Function Coverage (\%)} \\\cmidrule(lr){2-4}\cmidrule(lr){5-7}
            & Seeds &\afl  &\tool   & Seeds &\afl   &\tool\\\hline
\libplist   & 33.3  & 50.8 & 68.9   & 27.5  & 32.6  & 40.8\\
\webkit     & 52.4  & 56.0 & 78.0   & 35.1  & 37.0  & 49.5\\
\jerry      & 81.3  & 84.0 & 88.2   & 76.0  & 77.1  & 78.2\\
\chakra      & 46.7  & 54.5 & 76.9   & 40.7  & 49.8  & 63.2\\
\hline
\end{tabular}
\end{table}

For line coverage, the initial seeds covered 33.3\% lines of \libplist, 52.4\% lines of \webkit, 81.3\% lines~of~\jerry and 46.7\% lines of \chakra. By fuzzing, \afl~respectively increased their~line coverage to 50.8\%, 56.0\%, 84.0\% and~54.5\%. On average, \afl further covered 7.9\% of the code. \tool improved the~line coverage to 68.9\%, 78.0\%, 88.2\% and 76.9\%, respectively; and it further covered 24.6\%~of~the~code~on~average. Overall,~\tool~outperformed \afl by 16.7\%~in line coverage, because the grammar-awareness in \tool carries the fuzzing exploration towards the application execution stage.

On the other hand, for function coverage, the initial seeds covered 44.8\% functions on average, and \afl and \tool increased the function coverage to 49.1\% and 57.9\%, respectively. Generally, \tool outperformed \afl by 8.8\% in function coverage due to its grammar-awareness.

\vspace{2pt}
\noindent
\fbox{
\begin{minipage}{0.455\textwidth}
In summary, \tool can significantly improve the code coverage of coverage-based grey-box fuzzers (e.g., 16.7\% in line coverage and 8.8\% in function coverage).
\end{minipage}
}

\subsection{Effectiveness of Grammar-Aware Trimming (RQ3)}\label{sec_trim}

Table~\ref{tab:trim} compares the trimming ratio (i.e., the ratio~of~bytes trimmed from test inputs) and the grammar validity ratio~(i.e., the ratio of test inputs that are grammar-valid after trimming) using the built-in trimming in \afl and the tree-based trimming in \tool. Numerically, for \libplist, the built-in trimming in \afl trimmed out 21.7\%~of~bytes in \xml test inputs~on~average, while our tree-based~trimming trimmed out 11.7\%~on average. On the other hand, 74.1\% of test inputs after the built-in~trimming were grammar-valid, but 100\% of test inputs~after~our~tree-based trimming were grammar-valid and can be further used to conduct our grammar-aware mutation.

Similarly, the built-in trimming respectively trimmed out 10.6\%, 5.1\% and 12.7\% of bytes in \js test inputs~for \webkit, \jerry and \chakra, while our~tree-based~trimming respectively trimmed out 7.6\%, 4.7\% and 11.3\% for~\webkit, \jerry and \chakra. On the other~hand,~our~tree-based trimming increased the grammar validity ratio~for~\webkit, \jerry and \chakra from 86.4\%, 89.3\%~and~83.7\% to 100\%, which can facilitate our grammar-aware mutation by improving the chance of applying grammar-aware mutation (which~is~more effective in generating test inputs that can trigger new coverage as will be discussed in Section~\ref{sec_strategies}).

\begin{table}[!t]
\centering
\small
\caption{Comparison Results of Trimming Strategies}\label{tab:trim}
\def\arraystretch{1.25}
\setlength{\tabcolsep}{0.4em}
\vspace{-5pt}
\begin{tabular}{ccccc}
\hline
\multirow{2}{*}{Program} & \multicolumn{2}{c}{Trimming Ratio (\%)} & \multicolumn{2}{c}{Grammar Validity Ratio (\%)} \\ \cmidrule(lr){2-3}\cmidrule(lr){4-5}
             & Built-In  & Tree-Based   & Built-In & Tree-Based \\ \hline
\libplist & 21.7 & 11.7   &74.1 & 100 \\
\webkit & 10.6  & 7.6   &86.4 & 100 \\
\jerry & 5.1  & 4.7  & 89.3 & 100 \\
\chakra & 12.7  & 11.3  & 83.7 & 100 \\
\hline
\end{tabular}
\end{table}

\vspace{2pt}
\noindent
\fbox{
\begin{minipage}{0.455\textwidth}
In summary, although with a relatively low trimming~ratio, our grammar-aware trimming strategy can significantly~improve the grammar validity ratio for the test inputs after trimming, which facilitates our grammar-aware mutation.
\end{minipage}
}

\subsection{Effectiveness of Grammar-Aware Mutation (RQ4)}\label{sec_strategies}

\begin{figure*}[!t]
\centering
\includegraphics[width=0.98\textwidth]{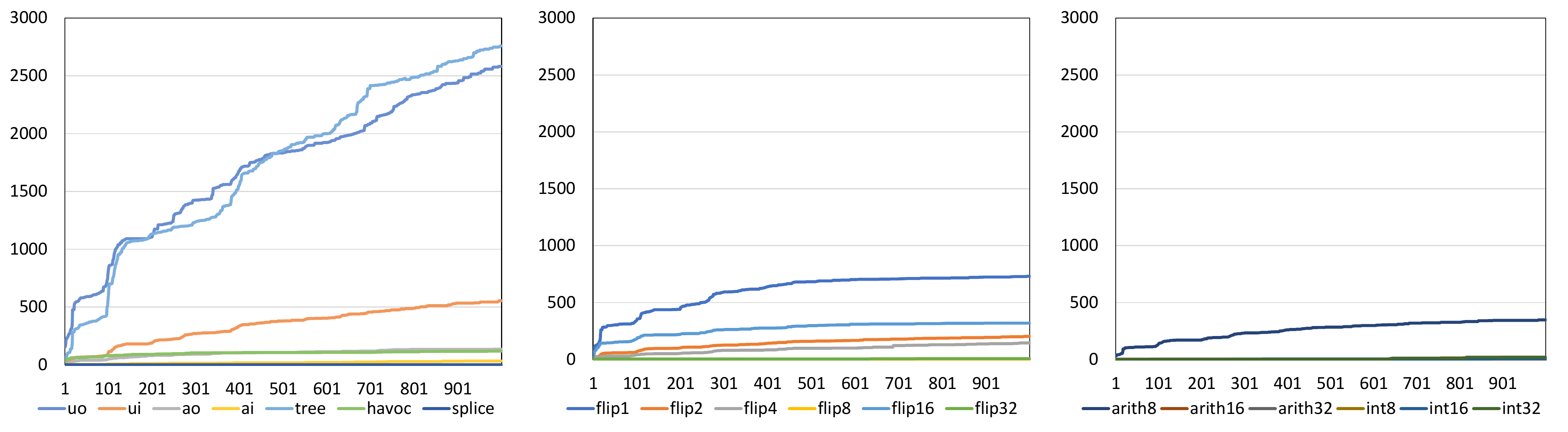}
\caption{The Effectiveness of Different Mutation Strategies in Producing Test Inputs that Trigger New Coverage}
\label{fig:mutator_stats}
\end{figure*}

\begin{figure*}[!t]
\centering
\includegraphics[width=0.98\textwidth]{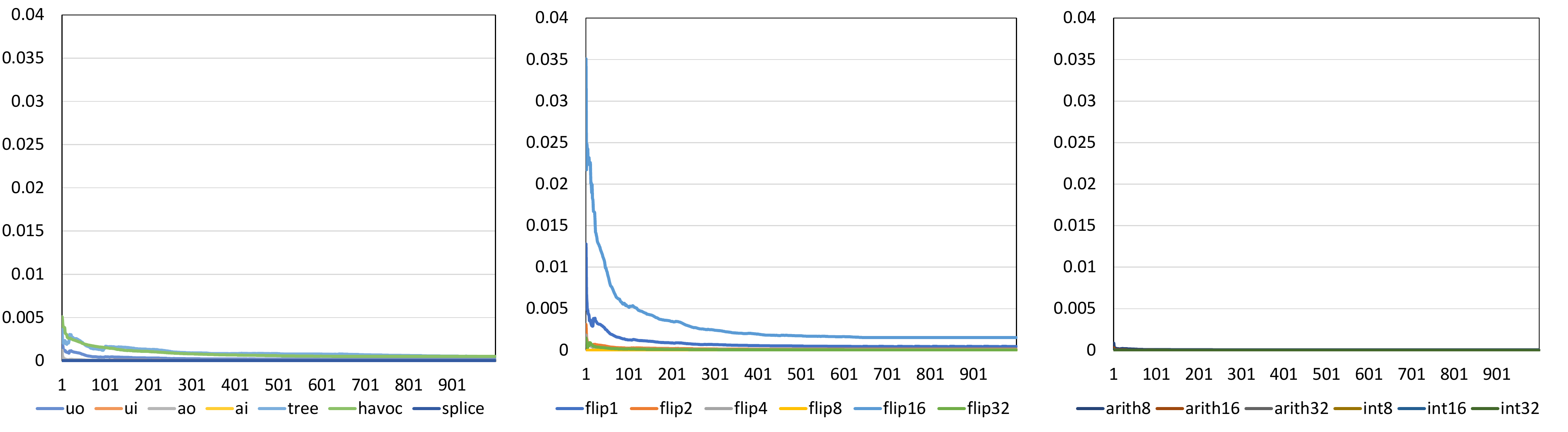}
\caption{The Efficiency of Different Mutation Strategies in Producing Test Inputs that Trigger New Coverage}
\label{fig:mutator_ratio}
\end{figure*}

To evaluate the effectiveness of our grammar-aware~mutation strategies, we compared them with those built-in mutation~strategies of \afl~\cite{aflmutation}, which include bit flips ({\it flip1}/{\it flip2}/{\it flip4}~--~one/two/four bit(s) flips), byte flips ({\it flip8}/{\it flip16}/{\it flip32}~--~one/two/four byte(s) flips), arithmetics ({\it arith8}/{\it arith16}/{\it arith32}~--~subtracting or adding small integers to 8-/16-/32-bit values), value overwrite ({\it interest8}/{\it interest16}/{\it interest32} -- setting ``interesting''~8-/16-/32-bit values to 8-/16-/32-bit values), {\it havoc} (random~application~of bit flips, byte flips, arithmetics, and value overwrite), and {\it splice} (splicing together two random test inputs from the~queue,~and then applying havoc). For the ease of presentation, our enhanced dictionary-based mutation strategy is referred as {\it ui} (insertion of user-supplied tokens), {\it uo} (overwrite with user-supplied tokens), {\it ai}~(insertion~of automatically extracted~tokens),~and~{\it ao}~(overwrite with automatically extracted tokens); and our tree-based mutation strategy is referred as {\it tree}.

Fig.~\ref{fig:mutator_stats} shows the number of interesting test inputs (i.e.,~triggering new coverage) discovered by different mutation strategies~as we fuzzed \webkit. For space limit, we~omit~the~similar~results for \libplist, \jerry and \chakra. The $x$-axis~denotes~the number of test inputs that \tool sequentially took from the queue and processed, and the $y$-axis denotes the corresponding number of interesting test inputs produced by different mutation strategies. As the process of different test inputs~often~takes~different time, we do not use time to represent the $x$-axis. Besides, for clarity, Fig.~\ref{fig:mutator_stats} omits the results when all the mutation strategies become~ineffective~in~continuously~producing~interesting test inputs (i.e., when the curves in Fig.~\ref{fig:mutator_stats} change gently).

The results vary across different seeds. Even with seeds~fixed, the results may also vary across different runs~due~to~the random nature of some mutation strategies (i.e., {\it havoc}, {\it splice} and {\it tree}).  However, the trend remains the same across runs, and we only discuss the trend which holds across runs. In the beginning, bit and byte flips take a leading position in producing interesting test inputs. The reasons are that~i)~bit~and~byte~flips~often destroy~the~input~structure, and trigger previously unseen error handling paths; and ii) bit and byte flips are the first mutation strategy to be sequentially applied, thus having the opportunity to first trigger the new coverage that could also be triggered by other mutation strategies. Gradually,~the~number~of~interesting~test~inputs~generated by our grammar-aware mutation strategies outperform other mutation strategies. Specifically, {\it tree} and {\it uo} significantly outperform other mutation strategies. These results indicate~that grammar-aware mutation strategies are effective in producing interesting test inputs.


Besides, we also explore the efficiency of different~mutation strategies in producing interesting test inputs. The results are shown in Fig.~\ref{fig:mutator_ratio}, where the $x$-axis is the same to Fig.~\ref{fig:mutator_stats}~and the $y$-axis denotes the ratio of interesting test inputs to the total number of generated test inputs. Surprisingly,~all~the~mutation strategies are very inefficient in producing interesting test inputs, i.e., only two of the 1000 mutated test inputs can trigger new coverage. Thus, a huge amount of fuzzing efforts are wasted in mutating and executed test inputs. Therefore, adaptive mutation rather than exhaustive mutation should be designed to smartly apply mutation strategies.

\begin{figure*}[!t]
\centering
\begin{subfigure}[b]{0.5\textwidth}
\centering
\includegraphics[width=0.8\textwidth]{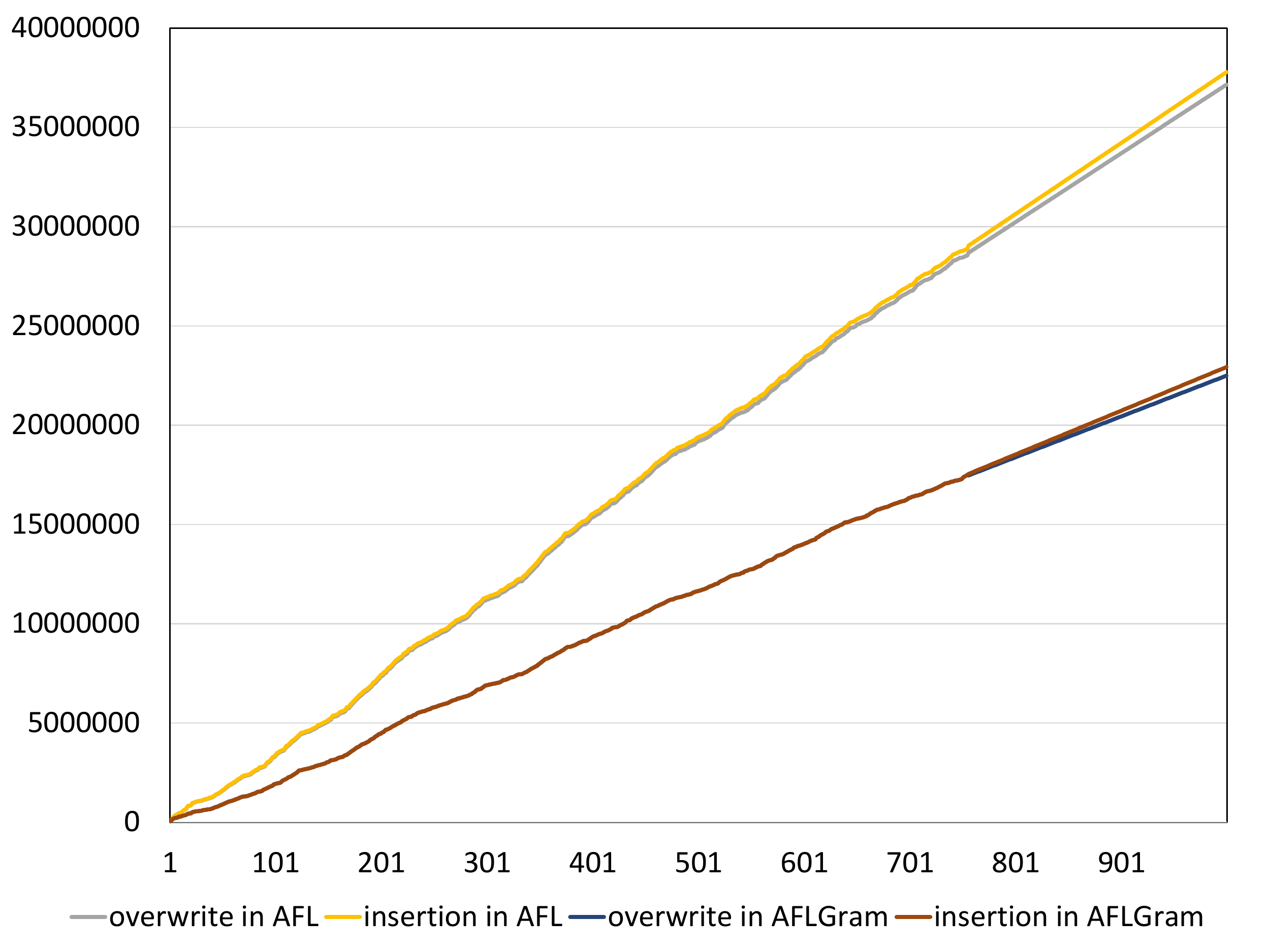}
\caption{The Number of Mutation Applications}\label{fig:dictionary_times}
\end{subfigure}%
~
\begin{subfigure}[b]{0.5\textwidth}
\centering
\includegraphics[width=0.8\textwidth]{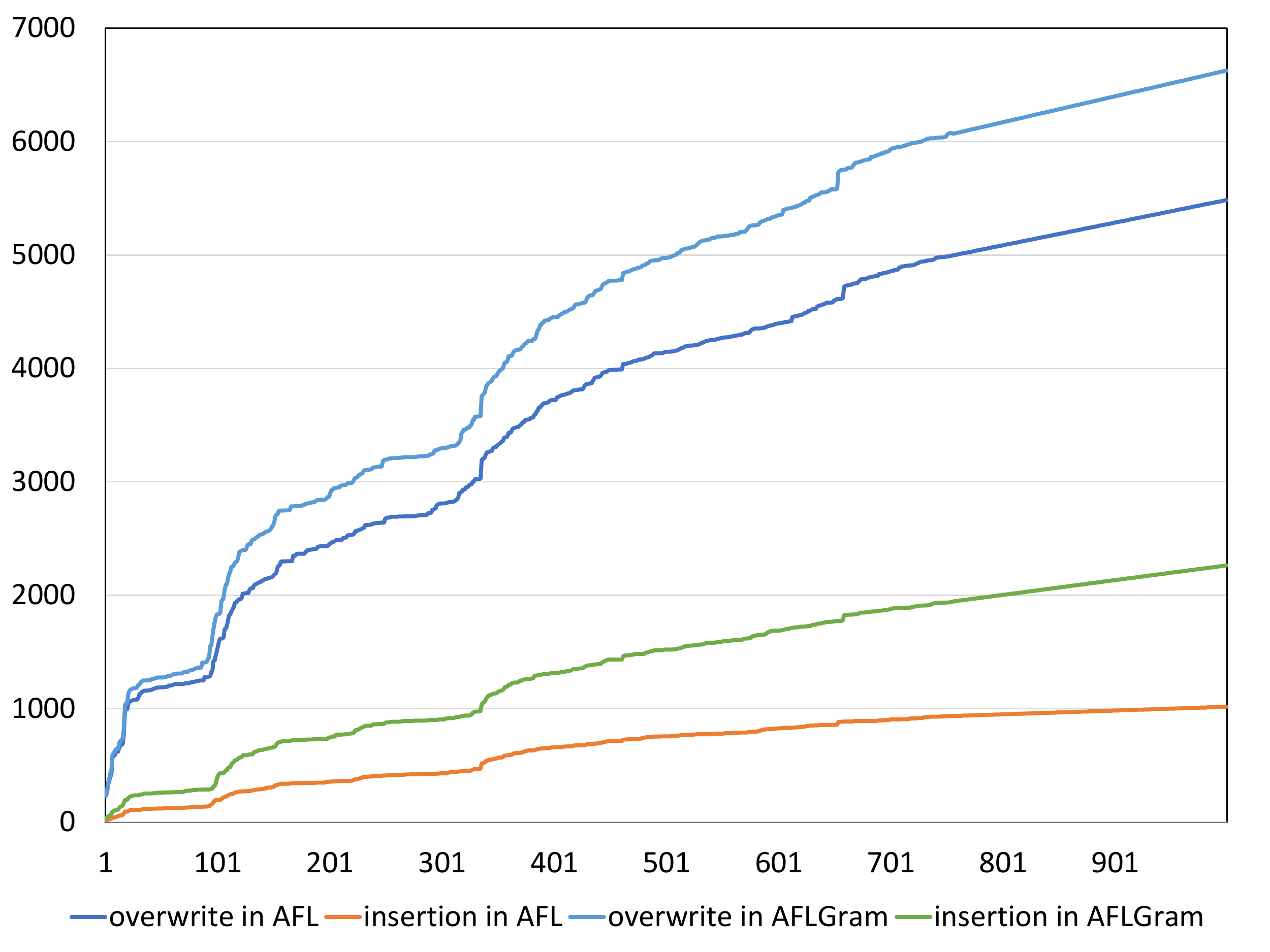}
\caption{The Effectiveness in Producing Test Inputs Triggering New Coverage}\label{fig:dictionary_new}
\end{subfigure}
\caption{Comparison Results of Dictionary-Based Mutations}\label{fig:dictionary}
\end{figure*}

Moreover, to evaluate our enhancement to dictionary-based mutation, we compared the dictionary overwrite and insertion~in \afl with those in \tool. The results are reported in Fig.~\ref{fig:dictionary}, where the $x$-axis is the same to Fig.~\ref{fig:mutator_stats}, and the $y$-axis in Fig.~\ref{fig:dictionary_times} and Fig.~\ref{fig:dictionary_new}  represent the number of times each mutation is applied and the number of interesting test inputs generated. We can see that our enhanced dictionary-based mutation greatly decreases the number of mutation applications by half, while still generating significantly more interesting test inputs.

\vspace{2pt}
\noindent
\fbox{
\begin{minipage}{0.455\textwidth}
In summary, our grammar-aware mutation strategies~are~effective in generating test inputs that can trigger new~coverage, compared to the built-in mutation strategies in \afl. However, the efficiency of all mutation strategies need to be improved.
\end{minipage}
}

\begin{figure*}[!t]
\centering
\begin{subfigure}[b]{0.45\textwidth}
\centering
\includegraphics[width=0.8\textwidth]{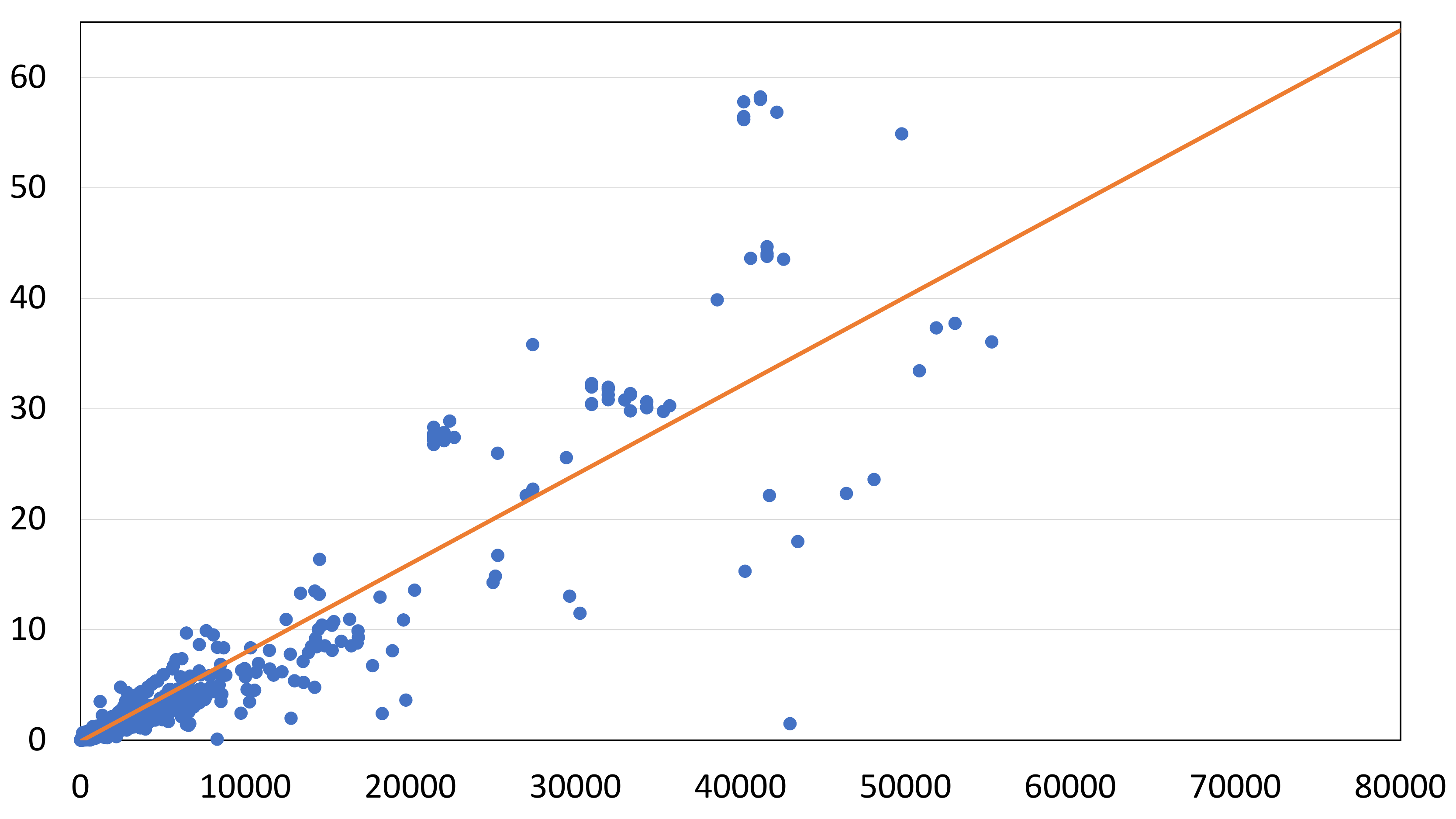}
\caption{\js Test Inputs}
\label{fig:parse_plus_traverse_time_js}
\end{subfigure}%
~
\begin{subfigure}[b]{0.45\textwidth}
\centering
\includegraphics[width=0.8\textwidth]{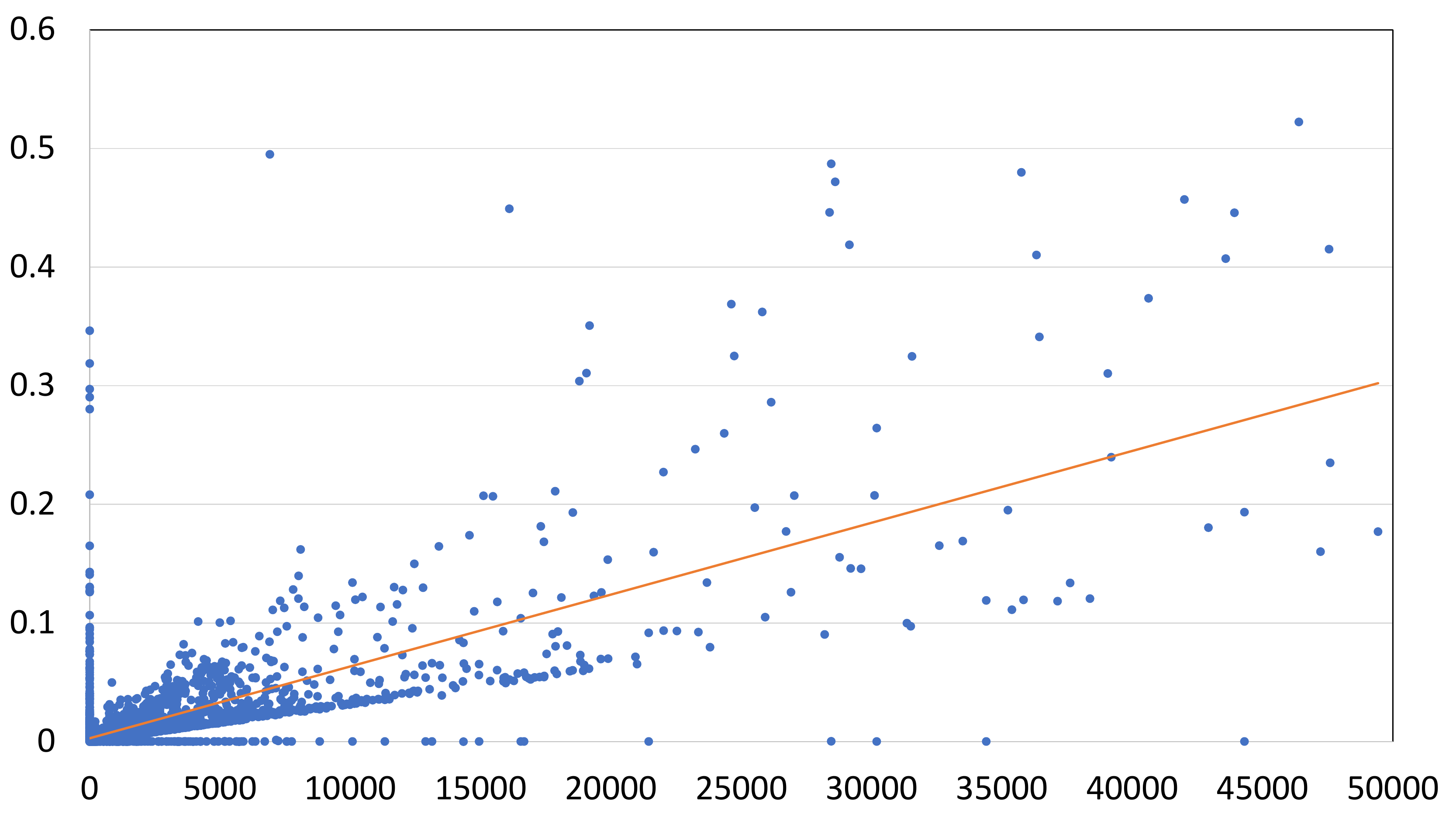}
\caption{\xml Test Inputs}
\label{fig:parse_plus_traverse_time_xml}
\end{subfigure}
\caption{The Time to Read, Parse and Traverse Test Inputs with Respect to Different Size}
\end{figure*}

\subsection{Performance Overhead (RQ5)}\label{sec_performance}
The fuzzing process of a test input includes three major steps: parsing, mutation and execution. Among them, the parsing~step is one-off for each test input, followed by a large number of mutations and executions. In Fig.~\ref{fig:parse_plus_traverse_time_js} and~\ref{fig:parse_plus_traverse_time_xml}, we show~the parsing time of \js and \xml test inputs in seconds~(the $y$-axis) with respect~to the size of test input files in bytes (the $x$-axis). Without loss of generality, we only report the results for the initial test inputs (see the last column in Table~\ref{tab:languages}). In detail, the parsing time includes the time to read, parse~and~traverse~a test input file. Generally, the parsing time is linearly correlated to the size of test input files. Most of the \js test inputs' size is under 10 KB and their parsing time is under 10 seconds, and the parsing time of \xml test inputs is under 0.045 seconds. Notice that the parser generated using ANLTR is not optimized for the performance. We may reduce the execution time further by improving the parser's implementation.

\begin{table}[!t]
\centering
\small
\caption{Performance Overhead on Target Programs}\label{tab:performance}
\def\arraystretch{1.25}
\setlength{\tabcolsep}{0.4em}
\vspace{-5pt}
\begin{tabular}{ccc}
\hline
Program & Tree-Based Mutation (ms) & Execution (ms) \\\hline
\libplist & 0.63 & 0.39 \\
\webkit & 5.65 & 12.50 \\
\jerry & 5.65 & 3.57 \\
\chakra & 5.65 & 20.00 \\
\hline
\end{tabular}
\end{table}

Apart from the parsing time, the major performance overhead \tool imposes on mutation and execution is caused by~our tree-based mutation. Table~\ref{tab:performance} reports the overhead of applying tree-based mutation (in the second column)~as~well~as~the~corresponding overhead of executing the mutated test input (in~the third column). For small projects like \libplist, it is very fast to perform tree-based mutation and execution, i.e., the mutation took 0.63 ms and the execution took 0.39 ms on average. For large projects such as \webkit, \jerry and \chakra, the execution took much more time; e.g., executing a \js input on \chakra took 20.00 ms, while the mutation took 5.65 ms on average.  Considering the improvements to bug-finding capability and code coverage, the performance overhead introduced by \tool is acceptable.

\vspace{2pt}
\noindent
\fbox{
\begin{minipage}{0.455\textwidth}
In summary, \tool introduces additional overhead due to our grammar-aware tree-based mutation strategy.~However, such overhead is still acceptable considering~the~improved bug-finding capability and code coverage.
\end{minipage}
}

\subsection{Case Study}

The \js code fragment in Fig.~\ref{fig:case_study_sample} gives a representative test input that was generated~by \tool and~triggered an integer overflow vulnerability in \webkit, assigned CVE-2017-7xxx. 
In particular, this vulnerability is triggered because the method {\tt setInput}  in class {\tt RegExpCachedResult} forgets to reify the {\tt leftContext} and {\tt rightContext}. As a result, when later \webkit attempts to reify them, it will end up using indices into an old input string to create a substring of a new input string. For the test input in Fig.~\ref{fig:case_study_sample}, \webkit tried to get a substring through {\tt jsSubstring}, whose length is 1 (i.e., length of ``a'') - 2 (i.e., {\tt m\_result.end} of ``ss'') = -1, as shown in Fig.~\ref{fig:case_study_code}, which is a very large number when treated as positive. Thus,~an integer overflow vulnerability is caused.


The test input in Fig.~\ref{fig:case_study_sample} was actually simplified from~a~large test input for the ease of presentation. It was generated~by~applying our tree-based mutation on the two test inputs in Fig.~\ref{fig:case_study_father} and Fig.~\ref{fig:case_study_mother}. This proof-of-concept was not generated through~one mutation, but was generated after several times of mutations. The intermediate test inputs that triggered new coverage were kept and added to the queue for further mutations. Eventually,~it evolved into the proof-of-concept. This vulnerability was not triggered by \afl. This indicates that AFL's built-in mutation strategies is not effective~in~fuzzing~programs~that~process~structured inputs, where our tree-based mutation becomes effective.

\begin{figure}[!t]
\scriptsize
\begin{lstlisting}[tabsize=2,frame=single,language=java,basicstyle=\ttfamily]
var str="ss";
var re=str.replace(/\b\w+\b/g);
RegExp.input="a";
RegExp.rightContext;
\end{lstlisting}
\caption{A Proof-of-Concept of CVE-2017-7xxx}
\label{fig:case_study_sample}
\end{figure}

\begin{figure}[!t]
\tiny
\begin{lstlisting}[frame=single,language=c,basicstyle=\ttfamily]
JSString* RegExpCachedResult::rightContext(ExecState* exec, JSObject* owner)
{
  // Make sure we're reified.
  lastResult(exec, owner);
  if (!m_reifiedRightContext) {
   unsigned length = m_reifiedInput->length();
   m_reifiedRightContext.set(exec->vm(), owner, m_result.end != length ?
      jsSubstring(exec, m_reifiedInput.get(), m_result.end, length - m_result.end)
      : jsEmptyString(exec));
  }
  return m_reifiedRightContext.get();
}
\end{lstlisting}
\caption{The Vulnerable Code Fragment for CVE-2017-7xxx }
\label{fig:case_study_code}
\end{figure}

\begin{figure}[!t]
\scriptsize
\begin{lstlisting}[tabsize=2,frame=single,language=java,basicstyle=\ttfamily]
...
var str = "ss"
var re=str.replace(/\b\w+\b/g);
...
\end{lstlisting}
\caption{Source Test Input to Trigger CVE-2017-7xxx}
\label{fig:case_study_father}
\end{figure}

\begin{figure}[!t]
\scriptsize
\begin{lstlisting}[tabsize=2,frame=single,language=java,basicstyle=\ttfamily]
...
write('RegExp.input: ' + RegExp.input);
...
write('RegExp.rightContext: ' + RegExp.rightContext);
...
\end{lstlisting}
\caption{Source Test Input to Trigger CVE-2017-7xxx}
\label{fig:case_study_mother}
\end{figure}

\subsection{Discussion}

One threat to the validity of our evaluation is that we did~not evaluate \tool on standardized data sets such as LAVA~\cite{dolan2016lava} and CGC~\cite{cgc}. However, many of the programs in these data~sets process unstructured inputs, or are difficult to come up with a grammar. Therefore, we did not use them. Instead, we used~four real-life programs, whose evaluation results are representative.

Another threat is that we did not empirically compare~\tool with LangFuzz~\cite{holler2012fuzzing} and IFuzzer~\cite{Veggalam2016}, two general-purpose grammar-aware mutation-based fuzzers. LangFuzz~is~not publicly available, and IFuzzer lacks sufficient documentation~to~set up. Instead, we compared \tool with jsfunfuzz, a successful grammar-aware generation-based fuzzer for \js engines.

One limitation of \tool is that it needs a grammar, which limits the applicability to only publicly documented formats that have specified grammars. Therefore, \tool may~have~trouble finding proprietary grammars or undocumented extensions~to standard grammars. However, several automatic grammar~inference techniques~\cite{viide2008experiences, hoschele2016mining, bastani2017synthesizing, Godefroid2017LML} have been proposed, we plan~to integrate such techniques to have a wider applicability.




\section{Related Work}\label{sec:related}

Instead of listing all related work, we focus our~discussion~on the most relevant fuzzing work in five aspects:~guided~mutation, grammar-based mutation, block-based generation, grammar-based generation, and fuzzing boosting.

\textbf{Guided Mutation.} Mutation-based fuzzing was proposed~to generate test inputs by randomly mutating well-formed test~inputs~\cite{Miller1990}.~Then, a large body of work has been developed~to~use heuristics~to guide mutation. \afl~\cite{afl}, Steelix~\cite{Li2017},~FairFuzz \cite{Lemieux2018S}~and CollAFL~\cite{Gan2018ZQTLPC} use coverage to achieve the guidance, and SlowFuzz~\cite{Petsios2017SAD} and PerfFuzz~\cite{Lemieux2018PAG} further use resource usage to realize the guidance. BuzzFuzz~\cite{ganesh2009taint}, Vuzzer~\cite{rawat2017vuzzer}~and~Angora \cite{Chen2018C} leverage taint analysis to identify those interesting bytes~for mutation. SAGE~\cite{Godefroid2008, Godefroid2012}, Babi\'{c} et al.~\cite{Babic2011},~Pham et al.~\cite{pham2016model} and Badger~\cite{Noller2018BCA} leverage symbolic execution to facilitate fuzzing. Dowser~\cite{Haller2013}, TaintScope~\cite{Wang2010} and BORG~\cite{Neugschwandtner2015} integrate taint analysis with symbolic execution to guide fuzzing.~Driller \cite{stephens2016driller} combines fuzzing and concolic execution to discover deep~bugs. Karg{\'e}n and Shahmehri~\cite{kargen2015turning} perform mutations on the~machine code of the generating programs~instead~of~directly~on~a~test~input in order to leverage the information about the input format encoded in the generating programs. In summary, these fuzzing techniques target programs that process~compact or unstructured inputs, which become less effective for programs that process structured inputs. Complementary~to~them, \tool~can~effectively fuzz programs that process structured inputs.

It is worth mentioning that application-specific fuzzers have been attracting great interests, e.g., compiler fuzzing~\cite{Chen2013TCF, le2014compiler, Lidbury2015MCF, Sun2016FCB, Le2015FDC, cummins2018compiler}, kernel fuzzing~\cite{schumilo2017kafl, Han2017IIM, Corina2017DIA}, IoT (Internet~of~Things) fuzzing~\cite{chen2018ndss}, OS fuzzing~\cite{Pailoor2018AJ}, smart contract fuzzing~\cite{Jiang2018LC}, GUI testing~\cite{Su2017GSM}, and deep learning system testing~\cite{Ma2018DMT}. It is interesting to investigate how to extend our general-purpose fuzzer (e.g., by designing new mutation operators or feedback mechanisms) to be effective in fuzzing specific applications.


\textbf{Grammar-Based Mutation.} Several~techniques have been proposed to perform mutations based on~grammar.~MongoDB's fuzzer \cite{guo2017mongodb} wreaks controlled havoc on the AST~of~a~\js test input. While our tree-based mutation~is similar, \tool conducts the mutations in an incremental way by keeping~those interesting intermediate test inputs for further fuzzing. Similarly, \si{\micro}4SQLi~\cite{appelt2014automated} applies a set of mutation operators on~valid~SQLs~to generate syntactically correct and executable SQLs that can~reveal SQL vulnerabilities. However, both MongoDB and \si{\micro}4SQLi are specifically designed for \js or SQL, and hence they may not work for other structured inputs. \tool~is general for other structured inputs as long as their grammar is available.


LangFuzz~\cite{holler2012fuzzing} uses a grammar to separate previously failing test input to code fragments and save them into~a~fragment pool. Then, some code fragments of a test~input~are~mutated by replacing them with the same~type~of~code~fragments~in~the pool. Similarly, IFuzzer~\cite{Veggalam2016} uses the grammar to extract~code~fragments from test inputs and recomposes them in an evolutionary~way. Different~from~these two~blackbox fuzzers, \tool~brings grammar-awareness into coverage-based greybox fuzzers. 


\textbf{Block-Based Generation.} As some bytes in a test input~are used collectively~as~a~single value in the program, they~should be considered together as a block~during fuzzing. Following this observation, TestMiner~\cite{dellasaying} first~mines literals from~a~corpus~of test~inputs and then queries the mined data for values suitable for a given method under test. These predicted values are then used as test inputs during test generation.~It~is~not~clear whether it works well for highly-structured inputs such as JavaScript~as they experimented with simple formats such as IBAN, SQL, E-mail and Network address. Spike~\cite{spike} and Peach \cite{eddington2011peach} use input models, specifying~the format of data chunks and integrity constraints, to regard test inputs as blocks of data,~and leverage mutations~to~generate~new~test~inputs. While being effective in fuzzing programs that process weakly-structured inputs (e.g., images and protocols), these approaches become less effective for highly-structured inputs (e.g., \js).~Complementary to them, \tool is designed for such highly-structured inputs.


\textbf{Grammar-Based Generation.} Another line of work is to use the grammar to directly~generate test inputs. mangleme~\cite{mangleme} is an automated broken HTML generator and browser fuzzer. jsfunfuzz~\cite{ruderman2007introducing} uses specific knowledge about past and common vulnerabilities and hard-coded rules to generate new test inputs. Dewey et~al.~\cite{Dewey2014} propose to use constraint logic programming for program generation. Valotta~\cite{valotta2012taking} uses his~domain~knowledge to manually build a fuzzer to test browsers. While being~effective in finding vulnerabilities, they all rely on some hard-coded or manually-specified rules to express semantic rules, which hinder their applications to a wider audience.


Godefroid et al.~\cite{godefroid2008grammar} apply symbolic execution~to~generate grammar-based constraints, and use grammar-based~constraint solver to generate test inputs. CSmith~\cite{yang2011finding} iteratively~and~randomly selects~one~production rule in the grammar~to~generate~C programs. Domato~\cite{domato} generates~test~inputs~from~scratch~given the grammars that specify~HTML/CSS structures and \js objects, properties and functions. Domato also fuzzed WebKit for three months; but none of our bugs were found~by Domato. This is a strong evidence that \tool has the characteristics that grammar-aware fuzzers without coverage feedback do not have. Skyfire \cite{wang2017skyfire} and~TreeFuzz \cite{patra2016learning} learn a probabilistic~model from the grammar~and~a~corpus of test inputs~to~generate test inputs. They are generation-based, while \tool~is~grammar-aware mutation-based, which incrementally utilizes the interesting behaviors embedded in previous interesting test inputs.


\textbf{Fuzzing Boosting.} Another thread of work focuses~on~improving the efficiency of fuzzing, e.g., seed selection~\cite{rebert2014optimizing},~seed scheduling~\cite{woo2013scheduling,bohmecoverage}, parameter tuning~\cite{Householder2012, cha2015program}, directed fuzzing \cite{bohme2017directed, hawkeye, Chen2018FVC} to reproduce crashes or assess potential bugs found by vulnerable code matching~\cite{Chandramohan2016BCC, Xu2017NNG}, and operating primitives \cite{Xu2017DNO}. These boosting~techniques are orthogonal to \tool. 

\section{Conclusions}\label{sec:conclusion}

In this paper, we propose a grammar-aware coverage-based greybox fuzzing approach, \tool, for programs~that process structured inputs. Specifically, we propose a grammar-aware trimming strategy and two grammar-aware mutation~strategies to effectively trim and mutate test inputs while keeping the input structure valid, quickly carrying the fuzzing exploration~into width and depth. Our~experimental study on several \xml~and \js engines has demonstrated that \tool improved code coverage and bug-finding capability over AFL. Moreover, \tool found 31 new bugs, among which 21 new~vulnerabilities were discovered and 16 CVEs were assigned.



{\footnotesize
\bibliographystyle{IEEEtranS}
\bibliography{ref}
}

\end{document}